\documentclass[useAMS,usenatbib]{mn2e}
\usepackage{epsf,graphicx,rotating,url}

\title[Near-IR radius-luminosity relationship]{The near-infrared
  radius-luminosity relationship for active galactic nuclei}

\author[H. Landt et al.]{Hermine Landt$^1$\thanks{E-mail: hlandt@unimelb.edu.au}\thanks{Visiting Astronomer at the Infrared Telescope Facility, which is operated by the University of Hawaii under Cooperative Agreement no. NNX-08AE38A with the National Aeronautics and Space Administration, Science Mission Directorate, Planetary Astronomy Program.}, Misty C. Bentz$^2$\footnotemark[2], Bradley M. Peterson$^{3,4}$, Martin Elvis$^5$\footnotemark[2], 
\newauthor
Martin J. Ward$^6$\footnotemark[2], Kirk T. Korista$^7$ and Margarita Karovska$^5$ \\ 
$^1$School of Physics, University of Melbourne, Parkville, VIC 3010, Australia \\ 
$^2$Department of Physics and Astronomy, Georgia State University, 709 One Park Place South, 
Atlanta, GA 30303, USA \\
$^3$Department of Astronomy, The Ohio State University, 
140 West 18th Avenue, Columbus, OH 43210, USA \\
$^4$Center for Cosmology and AstroParticle Physics, The Ohio State University, 
191 West Woodruff Avenue, Columbus, OH 43210, USA \\
$^5$Harvard-Smithsonian Center for Astrophysics, 60 Garden Street, 
Cambridge, MA 02138, USA \\
$^6$Department of Physics, University of Durham, South Road, Durham, DH1 3LE \\ 
$^7$Department of Physics, Western Michigan University, 
1903 W. Michigan Avenue, Kalamazoo, MI 49008, USA}

\begin{document}

\def\la{\mathrel{\hbox{\rlap{\hbox{\lower4pt\hbox{$\sim$}}}\hbox{$<$}}}}
\def\ga{\mathrel{\hbox{\rlap{\hbox{\lower4pt\hbox{$\sim$}}}\hbox{$>$}}}}

\font\sevenrm=cmr7
\def\OIII{[O~{\sevenrm III}]}
\def\FeII{Fe~{\sevenrm II}}
\def\SIII{[S~{\sevenrm III}]}
\def\HeI{He~{\sevenrm I}}
\def\HeII{He~{\sevenrm II}}
\def\NeV{[Ne~{\sevenrm V}]}
\def\OIV{[O~{\sevenrm IV}]}

\def\cloudy{{\sevenrm CLOUDY}}

\date{Accepted ~~. Received ~~; in original form ~~}

\pagerange{\pageref{firstpage}--\pageref{lastpage}} \pubyear{2011}

\maketitle

\label{firstpage}

\begin{abstract}

  Black hole masses for samples of active galactic nuclei (AGN) are
  currently estimated from single-epoch optical spectra. In
  particular, the size of the broad-line emitting region needed to
  compute the black hole mass is derived from the optical or
  ultraviolet continuum luminosity. Here we consider the relationship
  between the broad-line region size, $R$, and the near-infrared
  (near-IR) AGN continuum luminosity, $L$, as the near-IR continuum
  suffers less dust extinction than at shorter wavelengths and the
  prospects for separating the AGN continuum from host-galaxy
  starlight are better in the near-IR than in the optical. For a
  relationship of the form $R \propto L^{\alpha}$, we obtain for a
  sample of 14 reverberation-mapped AGN a best-fit slope of
  $\alpha=0.5\pm0.1$, which is consistent with the slope of the
  relationship in the optical band and with the value of 0.5 na\"ively
  expected from photoionisation theory. Black hole masses can then be
  estimated from the near-IR virial product, which is calculated using
  the strong and unblended Paschen broad emission lines (Pa$\alpha$ or
  Pa$\beta$).

\end{abstract}

\begin{keywords}
galaxies: active -- galaxies: nuclei -- infrared: galaxies -- quasars: general
\end{keywords}

\section{Introduction}

The discovery of tight correlations between a galaxy's central black
hole mass and the luminosity and velocity dispersion of its stellar
bulge \citep{Mag98, Geb00, Ferr00} is expected to strongly constrain
how galaxies form and grow over cosmic time. Therefore, much effort
goes in particular into measuring the rate of black hole growth
\citep[e.g.,][]{Yu02, Heck04, Kelly09}. Since this requires both large
samples of galaxies with easily obtainable black hole mass estimates
and sources that probe the highest redshifts, such studies rely
heavily on active galactic nuclei (AGN).

In AGN, black hole masses can be readily estimated from single-epoch
optical spectra. Assuming that the dynamics of the broad-emission line
gas seen in these sources is dominated by the gravitational force of
the black hole, one can use the virial theorem to calculate black hole
masses from only two observables, the velocity dispersion and radial
distance of the emitting gas from the central source
\citep[e.g.,][]{Pet04}. The velocity dispersion can be obtained from
the widths of the Balmer hydrogen lines (H$\beta$ or H$\alpha$) and
the continuum luminosity at (rest-frame) 5100~\AA~serves as a
surrogate for the size of the broad-line region (BLR). The latter
finds its justification in the work of \citet{Pet93}, \citet{Wan99}
and \citet{Kaspi00}, who showed that emission-line lags, and so
broad-emission line radii, derived from optical reverberation mapping
studies correlate with the optical luminosity (of the ionising
component) largely as expected from simple photoionisation
arguments. This correlation is now referred to as the
radius-luminosity ($R$--$L$) relationship.

We consider here the application of this technique in the
near-infrared (near-IR), which potentially affords some advantages
over the optical (and ultraviolet). First, the observed optical
continuum can be severely contaminated by host galaxy starlight if a
large slit is used (as is often the case in reverberation studies),
especially in low-redshift sources that have a weak AGN or sources
with a luminous stellar bulge, and so does not give directly the
ionising flux. High-resolution, deep optical images are then required
to estimate the host galaxy starlight enclosed in the aperture in
order to correct the optical spectra \citep{Bentz06a,
  Bentz09}. Secondly, the optical hydrogen broad-emission lines, in
particular H$\beta$, are strongly blended with other species, making
it difficult to reliably measure the line width. Thirdly, all optical
measures can suffer from dust extinction.

Here we present the radius--luminosity relationship in the near-IR,
which, in conjunction with the width of the strong and unblended
Paschen broad emission lines Pa$\alpha$ and Pa$\beta$, can be used to
estimate AGN black hole masses. In Section 2, we introduce the data
and present the near-IR $R$--$L$ relationship. Its application is
discussed in Section 3. In Section 4, we present our
conclusions. Throughout this paper we have assumed cosmological
parameters $H_0 = 70$ km s$^{-1}$ Mpc$^{-1}$, $\Omega_{\rm M}=0.3$,
and $\Omega_{\Lambda}=0.7$.

\section{The near-infrared $R$-$L$ relationship}

\begin{table}
\caption{\label{sample} 
Reverberation-Mapped Sample of AGN}
\begin{center}
\begin{tabular}{lcccc}
\hline
Object Name & $R_{\rm H\beta}$ & log $\nu L_{\rm 1\mu m}$ & log $\nu L_{\rm 1\mu m}^{\rm AGN}$ \\
& [lt-days] & [erg s$^{-1}$] & [erg s$^{-1}$] \\
(1) & (2) & (3) & (4) \\
\hline
3C273       & $307_{-91}^{+69}$       & 45.81$\pm$0.02 & 45.64$\pm$0.10 \\
Mrk876      &  40$\pm$15              & 44.77          & 44.55          \\
PG0844+349  & $32_{-13}^{+14}$        & 44.28$\pm$0.02 & 44.28$\pm$0.10 \\
Mrk110      & $\mathbf{26_{-6}^{+4}}$ & 43.69$\pm$0.08 & 43.66$\pm$0.05 \\ 
Mrk509      & $80_{-5}^{+6}$          & 44.05$\pm$0.03 & 44.07$\pm$0.04 \\ 
Ark120      & $\mathbf{40_{-6}^{+4}}$ & 44.26$\pm$0.06 & 43.70$\pm$0.10 \\ 
Mrk817      & $\mathbf{22_{-3}^{+2}}$ & 43.71$\pm$0.03 & 43.64$\pm$0.01 \\ 
Mrk290      &   9$\pm$1               & 43.43$\pm$0.04 & 43.22$\pm$0.12 \\ 
Mrk335      & $\mathbf{16_{-4}^{+3}}$ & 43.49$\pm$0.05 & 43.35$\pm$0.07 \\ 
Mrk79       & $\mathbf{15_{-5}^{+3}}$ & 43.30$\pm$0.04 & 43.20$\pm$0.05 \\ 
NGC5548     & $\mathbf{18.0\pm0.6}$   & 43.12$\pm$0.05 & 42.97$\pm$0.09 \\ 
NGC7469     & $4.5_{-0.8}^{+0.7}$     & 43.45          & 43.10          \\  
NGC4593     &   3.7$\pm$0.8           & 42.73$\pm$0.02 & 42.50$\pm$0.06 \\  
NGC4151     &   7$\pm$1               & 42.13$\pm$0.06 & 41.84$\pm$0.05 \\  
\hline
\end{tabular}

\parbox[]{7.6cm}{The columns are: (1) object name; (2) radius of the
  H$\beta$ broad-emission line region (in light-days) (H$\alpha$ for
  PG~0844$+$349); average integrated 1~$\mu$m continuum luminosity (4)
  as observed and (5) flux-calibrated relative to best-weather epoch
  and host galaxy-subtracted. Errors represent 1$\sigma$
  uncertainties. For sources with multiple radius measurements we
  list the weighted average (in boldface).}

\end{center}
\end{table}

\begin{figure}
\centerline{
\includegraphics[scale=0.45]{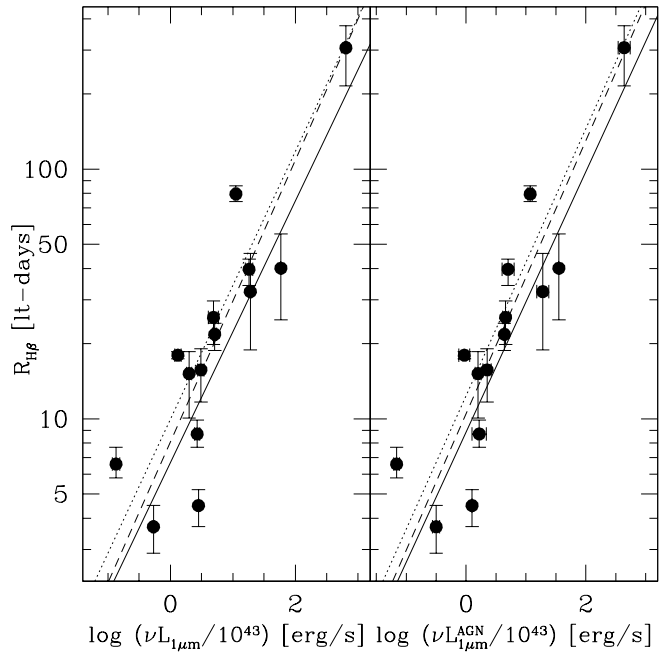}
}
\caption{\label{lag} The radius of the H$\beta$ broad-emission line
  region (in light-days) versus the average 1~$\mu$m continuum
  luminosity as observed (left panel) and flux-calibrated relative to
  best-weather epoch and host galaxy-subtracted (right panel). For
  sources with multiple radius measurements we plot the weighted
  average. The dashed, solid and dotted lines indicate the
  correlations obtained with the BCES, FITEXY and GaussFit routines,
  respectively.}
\end{figure}

We have recently shown \citep{L11a} that in broad-emission line AGN
the accretion disc spectrum, which is believed to be the main source
of ionising radiation, extends well into the near-IR and still
dominates the continuum at $\sim 1$~$\mu$m. Therefore, a single-epoch
near-IR spectrum can in principle be used to estimate the
broad-emission line radius.

In Fig. \ref{lag} we verify this conjecture by plotting the radius of
the H$\beta$ broad-emission line region ($R_{\rm H\beta}$) versus the
integrated 1~$\mu$m continuum luminosity for the reverberation-mapped
AGN in our sample (14 objects; see Table \ref{sample}). Values for the
radius of the H$\beta$ broad-line region are from \citet{Bentz09} and
for Mrk~290 from \citet{Denney10}. The near-IR measures are based on
data obtained with the SpeX spectrograph \citep{Ray03} at the NASA
Infrared Telescope Facility (IRTF) on Mauna Kea, Hawai'i. The
excellent atmospheric seeing at this observing site allowed us to use
a relatively narrow slit of $0.8''$, which excluded most of the host
galaxy starlight. The only exceptions were Mrk~590 and NGC~3227, two
reverberation-mapped AGN in our sample that were found to be in a very
low state with the continuum strongly dominated by host galaxy
emission \citep[see Fig. 6 in][]{L11a}. Therefore, they are omitted
from the present study. We observed sources on average twice within a
period of $\sim 3$ years \citep{L08a, L11a}. Here we use the mean
integrated 1~$\mu$m continuum luminosity and the error on the mean. We
consider both the observed values ($\nu L_{\rm 1\mu m}$) and the pure
AGN values ($\nu L_{1\mu m}^{\rm AGN}$). The latter were derived by
performing a relative flux-calibration using the \SIII~$\lambda 9531$
narrow emission line in the best-weather epoch and correcting for
host-galaxy starlight using the {\it Hubble Space Telescope (HST)}
images of \citet{Bentz06a, Bentz09} and for Mrk~290 a recent,
unpublished {\it HST} image (Bentz et al. in prep.). Further details
are provided by \citet{L11a}.

We have performed linear fits of the form \mbox{$\log R_{\rm H\beta} =
  K + \alpha\log(\nu L_{\nu}/10^{43})$} following the approach of
\citet{Bentz06a} and \citet{Bentz09}. In particular, we have used the
three fitting routines BCES \citep{Akr96b}, FITEXY \citep{recipes} and
GaussFit \citep{McA94} that can incorporate errors on both variables
and, except for GaussFit, allow us to account for intrinsic
scatter. Note that accounting for intrinsic scatter has the effect of
increasing the weight given to data points with the largest errors,
which is preferred if the intrinsic dispersion is larger than the
measurement errors \citep{Tre02}. Roughly half of our sample (6/14
sources) has multiple measurements of $R_{\rm H\beta}$ and, therefore,
we have considered the following two cases: (i) using the average
derived from all measurements for a particular source and weighted by
the mean of the positive and negative errors (i.e., weighted averages)
and (ii) using Monte-Carlo (MC) techniques to randomly sample $R_{\rm
  H\beta}$ from the individual values for each object. Two sources
have only one epoch of near-IR data and in these cases we have
assigned to the logarithmic luminosity the average error of the
sample. We obtain significant correlations in all cases. Table
\ref{fits} lists the results, which are shown for the case of the
weighted averages in Fig. \ref{lag}. Note that the near-IR continuum
luminosities and the optical broad-emission line radii are not
measured simultaneously, which is expected to increase the scatter in
their relationship.

All six cases for both total and AGN 1~$\mu$m continuum luminosity
give similar values for the slope and the intercept. The slope is in
most cases $\sim 0.5\pm0.1$ and always consistent with a value of
0.5. Simple photoionisation arguments suggest that a given emission
line will be produced at the same ionising {\it flux} in all AGN and
therefore $R \propto L^{1/2}$. Our results are similar to the best-fit
slope of $0.52^{+0.06}_{-0.07}$ found in the optical by
\citet{Bentz09}, who used host galaxy-subtracted fluxes of 34
reverberation-mapped AGN (more than twice our sample size).

\begin{table*}
\caption{\label{fits} 
Best-Fits for the Relation $\log R_{\rm H\beta} = K + \alpha\log(\nu L_{\nu}/10^{43})$}
\begin{tabular}{lcccccccccccc}
\hline
& \multicolumn{3}{c}{$\nu L_{\rm 1\mu m}$ (14 obj.)} & \multicolumn{3}{c}{$\nu L_{\rm 1\mu m}^{\rm AGN}$ (14 obj.)} 
& \multicolumn{3}{c}{$\nu L_{\rm 5100\AA}^{\rm AGN}$ (14 obj.)} & \multicolumn{3}{c}{$\nu L_{\rm 5100\AA}^{\rm AGN}$ (34 obj.)} \\
\hline
Type & K & $\alpha$ & S$^{\star}$ & K & $\alpha$ & S$^{\star}$ 
& K & $\alpha$ & S$^{\star}$ & K & $\alpha$ & S$^{\star}$ \\
\hline
\multicolumn{13}{c}{BCES} \\
\hline
Avg & 0.91$\pm$0.11 & 0.56$\pm$0.10 && 1.02$\pm$0.09 & 0.55$\pm$0.10 && 0.91$\pm$0.12 & 0.56$\pm$0.12 && 1.01$\pm$0.07 & 0.52$\pm$0.05 & \\
MC  & 0.90$\pm$0.13 & $0.56^{+0.12}_{-0.14}$ && $1.01^{+0.11}_{-0.12}$ & $0.55^{+0.11}_{-0.13}$ && 0.91$\pm$0.14 & 0.56$\pm$0.13 && 0.98$\pm$0.10 & 0.52$\pm$0.07 & \\
\hline
\multicolumn{13}{c}{FITEXY} \\
\hline
Avg & 0.82$\pm$0.10 & 0.52$\pm$0.10 & 53       & 0.95$\pm$0.08          & 0.52$\pm$0.08 & 48       & 0.84$\pm$0.09 & 0.53$\pm$0.10 & 50 & 0.91$\pm$0.06 & 0.54$\pm$0.05 & 40 \\
MC  & 0.81$\pm$0.11 & 0.53$\pm$0.11 & 52$\pm$2 & 0.93$\pm$0.09          & 0.52$\pm$0.09 & 48$\pm$2 & 0.82$\pm$0.11 & $0.53^{+0.10}_{-0.11}$ & 51$\pm$2 & $0.90^{+0.08}_{-0.07}$ & 0.54$\pm$0.06 & 41$\pm$1 \\
\hline
\multicolumn{13}{c}{GaussFit} \\
\hline
Avg & 1.00$\pm$0.10 & 0.54$\pm$0.10 &          & 1.09$\pm$0.08          & 0.54$\pm$0.09          & & 0.97$\pm$0.09 & 0.56$\pm$0.09 && 1.01$\pm$0.05 & 0.53$\pm$0.04 & \\
MC  & 0.95$\pm$0.14 & 0.56$\pm$0.12 &          & 1.07$\pm$0.11          & $0.54^{+0.09}_{-0.10}$ & & 0.92$\pm$0.12 & 0.58$\pm$0.10 && 0.96$\pm$0.08 & 0.56$\pm$0.05 & \\
\hline
\end{tabular}

\parbox[]{17.5cm}{$^{\star}$ Scatter calculated as the percentage of the $\log
  R_{H\beta}$ value that, when added in quadrature to the error value,
  gives $\chi^2_{\nu}=1$}

\end{table*}

\section{Discussion}

\begin{figure}
\centerline{
\includegraphics[scale=0.4]{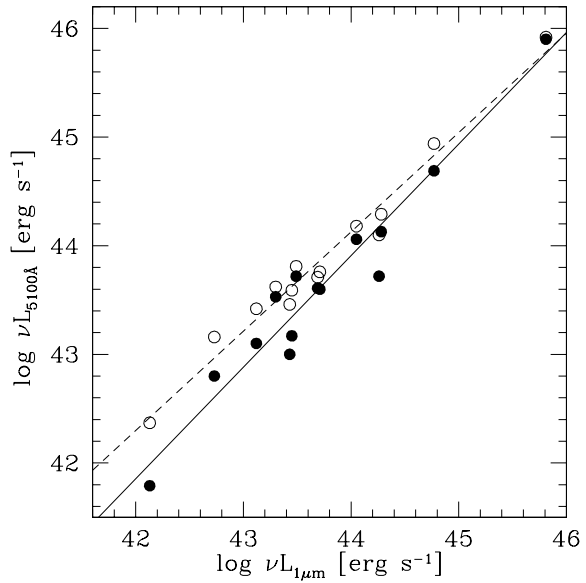}
}
\caption{\label{clum} The average integrated 5100~\AA ~continuum
  luminosity from \citet{Bentz09} versus the average integrated
  1~$\mu$m continuum luminosity. Open and filled circles indicate
  total and host galaxy-corrected optical values, respectively. The
  host galaxy correction to the optical luminosities has the effect of
  changing the observed correlation slope from $0.91\pm0.04$ (dashed
  line) to $1.03\pm0.07$ (solid line).}
\end{figure}

Given the importance of the $R$--$L$ relationship for AGN black hole
mass determinations, alternatives to the optical continuum luminosity
are already being explored, e.g., the X-ray luminosity and broad
H$\beta$ line luminosity \citep{Kaspi05, Greene10}, the \OIII~$\lambda
5007$~\AA~emission-line luminosity \citep{Greene10}, the Pa$\alpha$
and Pa$\beta$ emission-line luminosities \citep{Kim10}, and the
\NeV~$\lambda14.32~\mu$m and \OIV~$\lambda 25.89~\mu$m emission-line
luminosities \citep{Das08, Greene10}. The 1~$\mu$m continuum
luminosity is an efficient alternative that connects directly to the
spectrum of the ionising source.

Near-IR spectroscopy is now available at excellent observing sites
that regularly achieve subarcsecond seeing. This means that the host
galaxy flux contribution in the near-IR can be reduced to a
minimum. For example, a slit width (and seeing) of $0.8''$ in the
near-IR relative to one of $3''$, which is typical of most optical
telescope sites, leads to a reduction of the extraction aperture (and
so of the host galaxy flux contribution, assuming constant surface
brightness) by a factor of $\sim 14$. The case is exacerbated if we
consider the much larger apertures employed in optical reverberation
studies ($5-10''$).  Additionally, the seeing is improved at longer
wavelengths, thus reducing the extraction aperture further (e.g., by a
factor of $\sim 1.3$ at 1~$\mu$m relative to 5100~\AA). Adaptive
optics should make it possible to realise even greater gains in the
near-IR relative to the ground-based optical.

As our results show, the correlations for the total and AGN 1~$\mu$m
continuum luminosity are consistent with each other and so independent
of the host galaxy flux subtraction, although the slope is more often
$\sim 0.5$ and the scatter slightly smaller for the latter. Performing
the same fits for our sample (14 sources) using the optical data of
\citet{Bentz09} with the host galaxy flux subtracted gives results
that are more similar to the total rather than to the AGN near-IR
spectra (see Table \ref{fits}). However, if we consider their entire
optical sample (34 sources), we obtain results similar to the AGN
near-IR case. In this respect, we note that the intercepts of the
optical and near-IR $R$-$L$ relationships are similar, a result which
is not unexpected if the two AGN luminosities sample the canonical
accretion disc spectrum ($f_\nu \propto \nu^{1/3}$). In this case, the
difference in integrated luminosity would be $\sim 0.4$ dex, well
within the scatter of \mbox{$\sim 0.5$ dex} of the near-IR and optical
relationships for our sample (see Table \ref{fits}). Fig. \ref{clum}
further illustrates the efficacy of our approach to minimising the
host-galaxy contamination of the near-IR AGN continuum. While the
near-IR luminosities we measure correlate well with the measured
optical luminosities (open circles), with a slope of $0.91\pm0.04$,
they correlate even better once the host-galaxy contributions to the
optical luminosities have been removed. The slope of the relationship
between the optical AGN luminosities and the near-IR luminosities is
$1.03\pm0.07$. Clearly the narrow-slit near-IR luminosities are
dominated by the AGN component.

As we noted earlier, attenuation by dust is much reduced in the
near-IR compared to the optical, so the 1~$\mu$m continuum luminosity
may be particularly useful as an alternative for dust-obscured AGN. A
modest intrinsic obscuration of a hydrogen column density of $N_{\rm
  H} \sim 10^{22}$~cm$^{-2}$ results in a flux attenuation a factor of
$\sim 40$ smaller at 1~$\mu$m than at 5100~\AA~(in the
rest-frame). The case is even more extreme for Compton-thick AGN with
their expected hydrogen column densities in excess of $N_{\rm H} \ga
10^{24}$~cm$^{-2}$ \citep[e.g.,][]{Daddi07}.

As we have shown in \citet{L11a}, AGN black holes masses can be
estimated from the near-IR virial product based on the 1~$\mu$m
continuum luminosity and the width of the Pa$\beta$ (or Pa$\alpha$)
broad emission line because the widths of the broad Paschen lines are
well-correlated with those of the broad Balmer lines \citep{L08a}.
The advantages of using the near-IR instead of the optical virial
product are threefold. First, since Pa$\alpha$ and Pa$\beta$ are
observed to be unblended \citep{L08a}, their width can be reliably
measured. By contrast, the H$\beta$ broad emission line is generally
observed to be strongly blended with both \FeII~and
\HeII~$\lambda4686$ and often shows a ``red shelf'' most likely formed
by weak \FeII~multiplets and \HeI. Secondly, the AGN continuum around
1~$\mu$m is free from major contaminating components and may be easily
determined, unlike the optical, which can suffer from an
\FeII~pseudo-continuum. Finally, since the near-IR is much less
affected by dust extinction, it can potentially be applied to
dust-obscured AGN.

\section{Conclusions}

We have presented the near-IR radius--luminosity relationship. Using
near-IR spectra of 14 reverberation-mapped AGN obtained through a slit
small enough that it excludes most of the host galaxy starlight, we
have fit the relationship between the radius of the H$\beta$
broad-emission line region and the integrated 1~$\mu$m continuum
luminosity. The best-fit slope is in most cases $\sim 0.5\pm0.1$ and
always consistent with a value of 0.5, which is expected based on
simple photoionisation arguments. The near-IR $R$--$L$ relationship as
an alternative to the optical relationship is expected to be relevant
particularly for dust-obscured AGN. Black hole mass estimates can be
obtained from the near-IR virial product based on the integrated
1~$\mu$m continuum luminosity and the widths of the strong and
unblended Paschen broad emission lines Pa$\alpha$ or Pa$\beta$.

\section*{Acknowledgments}

We thank the anonymous referee for comments that helped us to improve
the paper. B. M. P. acknowledges support by the National Science
Foundation (NSF grant no. AST-1008882). This research has made use of
the NASA/IPAC Extragalactic Database (NED), which is operated by the
Jet Propulsion Laboratory, California Institute of Technology, under
contract with the National Aeronautics Space Administration.

\bibliography{/Users/herminelandt/references}

\bsp
\label{lastpage}

\end{document}